%% file: main.tex
\documentclass{IEEEcsmag}

\usepackage[colorlinks,urlcolor=blue,linkcolor=blue,citecolor=blue]{hyperref}
\expandafter\def\expandafter\UrlBreaks\expandafter{\UrlBreaks\do\/\do\*\do\-\do\~\do\'\do\"\do\-}
\usepackage{upmath,color}
\usepackage{xspace}

\jvol{XX}
\jnum{XX}
\paper{8}
\jmonth{Apr}
\jname{Publication Name}
\jtitle{Publication Title}
\pubyear{2025}

\newcommand{\figref}[1]{Figure~\ref{#1}\xspace}

\newcommand{\tabref}[1]{Table~\ref{#1}\xspace}

\newcommand{\topicfmt}[1]{\textit{#1}}
\newcommand{\enquote}[1]{``#1''}
\newcommand{\etal}[1]{et~al.}

\begin{document}

\sptitle{Article Type: Description}

\title{Understanding Prompt Management in GitHub Repositories: A Call for Best Practices}

\author{Hao Li}
\affil{Queen's University}

\author{Hicham Masri}
\affil{Queen's University}

\author{Filipe R. Cogo}
\affil{Huawei Technologies}

\author{Abdul Ali Bangash}
\affil{$L$ahore University of Management Sciences}

\author{Bram Adams}
\affil{Queen's University}

\author{Ahmed E. Hassan}
\affil{Queen's University}

\markboth{THEME/FEATURE/DEPARTMENT}{THEME/FEATURE/DEPARTMENT}

\begin{abstract}
\looseness-1
The rapid adoption of foundation models (e.g., large language models) has given rise to promptware, i.e., software built using natural language prompts. Effective management of prompts, such as organization and quality assurance, is essential yet challenging. In this study, we perform an empirical analysis of 24,800 open-source prompts from 92 GitHub repositories to investigate prompt management practices and quality attributes. 
Our findings reveal critical challenges such as considerable inconsistencies in prompt formatting, substantial internal and external prompt duplication, and frequent readability and spelling issues. 
Based on these findings, we provide actionable recommendations for developers to enhance the usability and maintainability of open-source prompts within the rapidly evolving promptware ecosystem.
\end{abstract}

\maketitle

\input{sections/1_introduction}
\input{sections/2_dataset}
\input{sections/3_patterns}
\input{sections/4_quality}
\input{sections/6_threats}
\input{sections/5_lessons}

\def\refname{REFERENCES}

\vspace*{-8pt}

\begin{IEEEbiography}{Hao Li}{\,} is a postdoctoral researcher at Queen’s University, Kingston, ON K7L 3N6, Canada. His research interests include software engineering for AI, AI for software engineering, and software package ecosystems. Li received his Ph.D. in Software Engineering and Intelligent Systems from the University of Alberta. Contact him at hao.li@queensu.ca.\vspace*{8pt}
\end{IEEEbiography}

\begin{IEEEbiography}{Hicham Masri}{\,} is a software and data engineer at DocuPet in Ontario, Canada. His research interests include machine learning, foundation models, and data-centric AI. Masri received his MSc in Computing from Queen’s University. Contact him at masri@queensu.ca.\vspace*{8pt}
\end{IEEEbiography}

\begin{IEEEbiography}{Filipe R. Cogo} {\,} is a software engineering researcher at Huawei Technologies Co., Kingston, ON, K7K 1B7, Canada. His research interests include machine learning and mining software repositories to investigate and propose automated solutions to technical and social problems in software engineering. Cogo received his Ph.D. in computer science from Queen’s University. Contact him at filipe.cogo@gmail.com.\vspace*{8pt}
\end{IEEEbiography}

\begin{IEEEbiography}{Abdul Ali Bangash}{\,} is a Tenure-track Assistant Professor at Lahore University of Management Sciences in Lahore, Pakistan. His research interests include using AI-based systems, mining software repositories and performance engineering to improve software and machine learning processes. Bangash received his PhD in Computer Science from the University of Alberta.  Contact him at abdulali@lums.edu.pk.\vspace*{8pt}
\end{IEEEbiography}

\begin{IEEEbiography}{Bram Adams} {\,} is a full professor at Queen’s University, Kingston, ON K7L 3N6, Canada. His research interests include software release engineering (pre- and post-AI) and mining software repositories. He is a Senior Member of IEEE. Contact him at bram.adams@queensu.ca.\vspace*{8pt}
\end{IEEEbiography}

\begin{IEEEbiography}{Ahmed E. Hassan} {\,} is the Natural Sciences and Engineering Research Council of Canada/Research in Motion Industrial Research chair in Software Engineering for Ultra Large Scale systems at Queen’s University, Kingston, ON K7L 3N6, Canada. He is a Fellow of the IEEE. Contact him at ahmed@cs.queensu.ca.
\end{IEEEbiography}

\end{document}

%% file: sections/1_introduction.tex
\chapteri{T}he popularization of foundation models~(FMs), such as GPT, Llama and DeepSeek, has given rise to \textit{promptware}, a new class of software built with natural language \emph{prompts}~\cite{hassan2024rethink}. Promptware democratizes software creation, allowing users with little or no training in coding or AI to build intelligent software applications. A prominent example is ChatGPT, a chat assistant that uses built-in prompts to define the assistant's behavior, tone, expertise, and formatting during conversations.

When building promptware, it is crucial to carefully manage the prompts used by the application, ensuring their proper storage, organization, versioning, and maintenance. While specialized \textit{prompt stores} such as PromptBase~\cite{promptbase} have emerged, they primarily function as commercial marketplaces for individual prompt assets rather than collaborative development platforms. Prior work has studied the evolution of prompts embedded directly into source code files (e.g., as Python strings)~\cite{tafreshipour2025promptingwildempiricalstudy}. However, no prior research has studied prompts as standalone software assets that are stored in dedicated files or investigated their management practices.

GitHub, as the leading collaboration platform for software development, has naturally emerged as a popular open-source choice for prompt management. However, as a Git-based platform, GitHub has fundamental limitations in managing prompt assets. First, in contrast to source code, prompts are unstructured or semi-structured at best. Second, there seems to be an impedance mismatch between prompts and GitHub's focus on source files as the unit of work and their respective lines as the unit of management. Finally, while GitHub offers a set of gatekeeping tools (e.g., GitHub Actions) to control the quality of the integrated source code, the same does not exist to ensure the quality of prompts. Despite these limitations, GitHub remains the de facto standard for open-source prompt management due to the lack of dedicated collaborative platforms that support version control and community contribution. As a result, the quality of the integrated prompts on GitHub still requires investigation.

In this article, we empirically study how developers manage open-source prompts on GitHub and evaluate the quality of these prompts. To ground our analysis in real-world evidence, we collect and analyze a dataset consisting of 24,800 prompts from 92 GitHub repositories. This dataset provides a detailed examination of current prompt management practices adopted by developers, highlighting both prevalent patterns and notable areas for improvement.

%% file: sections/2_dataset.tex
\section{Dataset Composition}

We construct our prompt dataset by mining repositories from GitHub and applying a topic modeling approach to uncover the primary themes within prompts, enabling an overview of the use cases and application domains.

\subsection{Mining Prompts from GitHub}

Using the GitHub API, we collect open-source prompts by searching GitHub repositories containing the keyword \enquote{prompts} in their name or description, created between December 2021 and December 2023.
This period captures the rise in prompt usage and sharing for building promptware, coinciding with major advances in public access to foundation models, notably the releases of GPT-3.5 (March 2022) and ChatGPT (November 2022)~\cite{openai2022chatgpt}.
We collect 19,103 repositories from the initial search and filter out 17,908 repositories with fewer than 10 stars, retaining 1,195 candidates. We restrict our selection to repositories that store prompts exclusively as standalone assets (e.g., TXT or Markdown) rather than source code. We manually filter out 911 repositories containing primarily programming languages such as Python and Java. This distinguishes our study from prior work~\cite{tafreshipour2025promptingwildempiricalstudy} that analyzes prompts embedded as string literals within application code, allowing us to investigate management practices specific to prompt files.

The data collection process yields a total of 284 initial repositories. To ensure relevance, we manually inspect each repository to confirm it contains at least one prompt, resulting in a refined set of 92 relevant repositories. From this set, we then manually extract prompts from smaller and less structured repositories while we develop custom scripts to automatically extract the prompts of larger and more structured repositories~(e.g., those storing prompts in CSV formats). Finally, we remove 720 non-English prompts from the collected data, resulting in a final dataset comprising 24,800 prompts from 92 repositories.

\subsection{Topic Analysis of the Prompt Dataset}


To provide a descriptive overview of the collected data, we follow the clustering approach and hyperparameters used by Zheng~et~al.~\cite{zheng2023lmsys} for topic modeling. 
First, we remove prompts that are either too short~(fewer than 32 characters) or excessively long~(greater than 1,536 characters). This filtering step removes approximately 2,530 prompts, leaving us with a final set of 22,270 prompts suitable for topic analysis. Next, we compute sentence embeddings for these prompts using the \enquote{all-mpnet-base-v2} model from SentenceTransformers~\cite{reimers2019sentencebert} and apply K-Means clustering with $k=20$ to group these embeddings into distinct thematic clusters, leveraging BERTopic~\cite{grootendorst2022bertopic}. We then select the 20 most representative prompts for each cluster by computing the cosine similarity between each prompt embedding and the corresponding cluster embedding, identifying prompts that best capture the central theme of each topic. Finally, we use GPT-4 to summarize the central topics covered by these representative prompts.

\begin{figure*}[t]
	\centering
	\includegraphics[width=\textwidth]{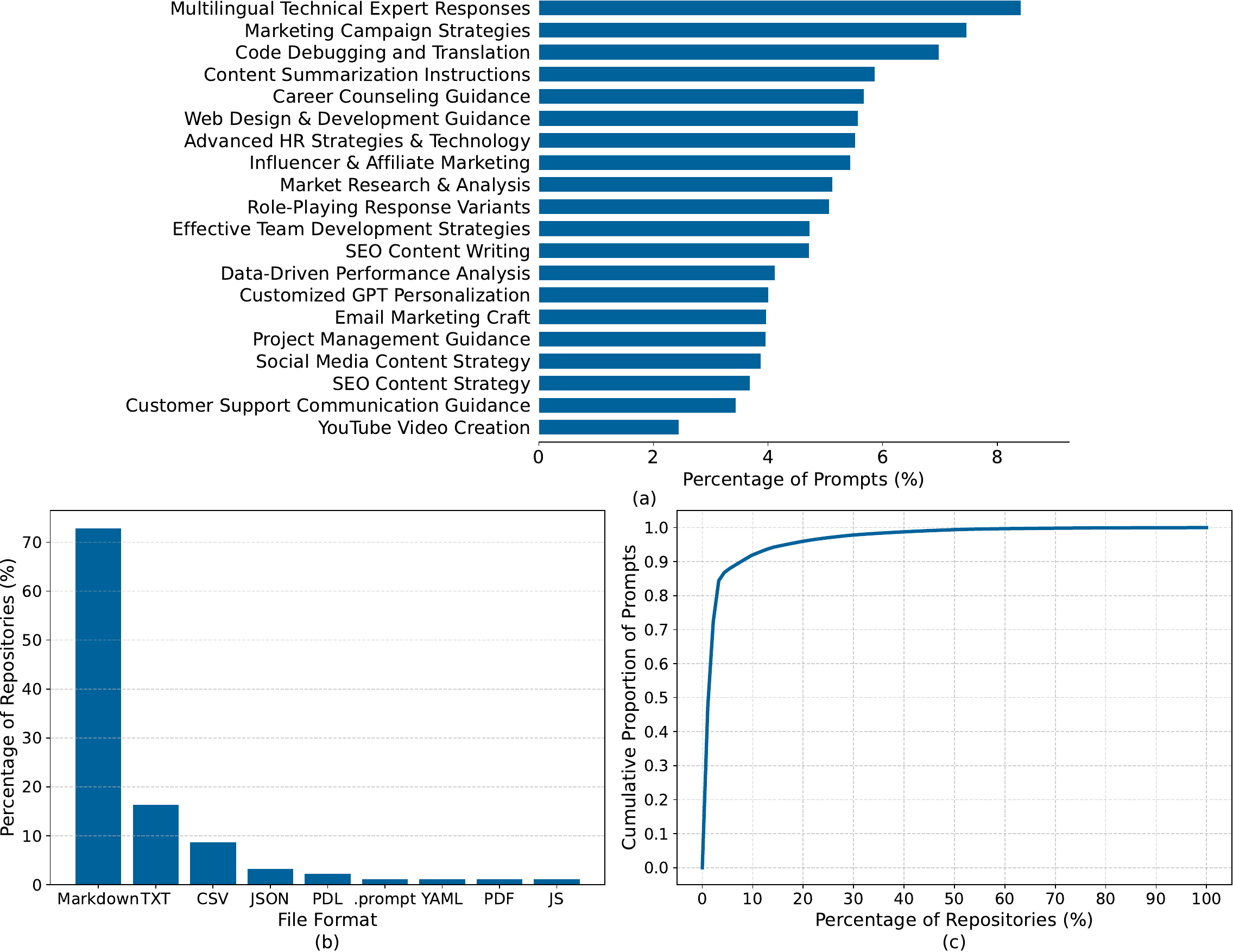}
	\caption{(a) Topic distribution of prompts in our dataset. (b) Frequency distribution of file formats used for storing prompts on GitHub. (c) Cumulative percentage of repositories plotted against the cumulative number of prompts.}
	\label{fig:format_and_num_prompts}
\end{figure*}

The distribution of the resulting topics is shown in \figref{fig:format_and_num_prompts} (a). Our analysis reveals that open-source prompts in GitHub prompt stores primarily concern marketing-related tasks, especially topics such as \topicfmt{Marketing Campaign Strategies}, \topicfmt{Market Research \& Analysis}, and \topicfmt{Email Marketing Craft}. An example prompt is ``Enhance the appeal of the \{ad copy\} by rewriting it to make it more persuasive.'' Other prominent prompt topics focus on content generation, such as \topicfmt{Content Summarization Instructions} and \topicfmt{SEO Content Writing}.

In addition, prompts related to \topicfmt{Code Debugging and Translation} emerge as the third-largest topic, accounting for around 7\% of all analyzed prompts. While coding tasks are less prevalent among prompt repositories than traditional code repositories, the topic of \topicfmt{Code Debugging and Translation} is notably widespread in 52.2\%~(48 out of 92) of all analyzed repositories. Similarly, prompts related to \topicfmt{Web Design \& Development Guidance} represent roughly 6\% of the dataset. These observations indicate a notable interest in leveraging promptware to support software engineering tasks.

\subsection{Categories of Prompt Repositories}

We manually examine the 92 repositories to gain an understanding of their varying objectives. Specifically, we categorize each repository based on its contents, stated purpose, and accompanying documentation~(e.g., README files and repository descriptions). The first and second authors independently classify the repositories, achieving an inter-rater reliability of Cohen's $\kappa=0.83$ (almost perfect agreement), and all disagreements are resolved through discussion.
We observe that they generally fall into three distinct objective-based categories: 

\begin{itemize}
\item[{\ieeeguilsinglright}] \textit{Prompt Collection.}
72.8\%~(67 out of 92) of the repositories store a large set of prompts that allow developers to reuse the prompts without reinventing the wheel. Usually, these repositories provide extensive collections of prompts without explicit authorship attribution or categorization. Only 7.5\%~(5 out of 67) of these prompt collections provide some form of authorship attribution, for example, including the author's name alongside the prompt. However, due to limited documentation and the absence of verification mechanisms, confirming the authenticity of these attributions remains challenging.

\item[{\ieeeguilsinglright}] \textit{Prompt Application.}
21.7\%~(20 out of 92) of the repositories contain prompts that are used for the development and maintenance of applications with a single, specialized purpose. For instance, an AI-driven tutoring application~\cite{mr_ranedeer} relies on a sophisticated single prompt~(approximately 1.7k words) to provide customized educational experiences for users with diverse learning objectives. 
The prompt acts as the source code of the application, and the application might use different prompts for different model sizes~(e.g., GPT-4o and GPT-4o-mini) and different model versions~(GPT-4o and GPT-5).

\item[{\ieeeguilsinglright}] \textit{Prompt Courseware.}
5.4\%~(5 out of 92) of the repositories serve as centralized educational resources or informational distributors. These repositories provide collections of research literature, prompt engineering guidance, best-practices documents, instructional tutorials, and other explanatory materials aimed at assisting users in prompt creation and optimization.

\end{itemize}

%% file: sections/3_patterns.tex
\section{Patterns and Chaos in Prompt Management}

Inefficient prompt management can lead to challenges for users attempting to discover prompts and for developers aiming to maintain the repositories. Therefore, we conduct an in-depth analysis of our collected dataset to identify prevailing patterns, organizational strategies, and potential pitfalls of current prompt management practices on GitHub.

To better understand prompt storage and organizational practices, we examine prompt storage formats~(e.g., TXT) and distinguish between repositories storing single-prompt files (where each file contains a single prompt) and multi-prompt files (where each file contains two or more prompts). In addition, we investigate the distribution of the number of prompts across the repositories and explore the issue of prompt duplication both within and across repositories, analogous to the code-duplication problems frequently observed in traditional software projects~\cite{rattam2013clone}. Beyond exact string matching, we assess semantic duplication via sentence embeddings (using ``all-mpnet-base-v2'') with a cosine similarity threshold of $>0.95$.

\subsection{Prompt Storage and Organization Practices}



\textbf{Markdown is the most popular prompt storage format (72.8\%), followed by TXT (16.3\%), despite TXT files being unstructured and potentially difficult to parse or reuse systematically.}
As shown in Figure~\ref{fig:format_and_num_prompts} (b), Markdown is used by 72.8\%~(67 out of 92) of analyzed repositories, which offers a lightly structured alternative to TXT files. Less prevalent storage formats include CSV~(8.7\%), JSON~(3.3\%), and several uncommon file formats~(each around 1\%), such as PDL, \enquote{.prompt}, JS, PDF, and YAML. Notably, the \enquote{.prompt} and PDL are specifically designed for prompt storage, but remain scarcely adopted. In addition, the majority of repositories show a clear preference for uniform file format usage, with 92.4\%~(85 out of 92) adopting only a single format for storing prompts. 
Prompt application repositories mainly use Markdown (12 out of 20) and TXT (6 out of 20), reflecting the need for simplicity and ease of use. Meanwhile, prompt courseware repositories predominantly rely on Markdown (4 out of 5), with 1 repository using PDF, likely due to the ease of reading and presentation these formats offer for instructional materials. Prompt collection repositories use a wider range of formats beyond Markdown (51 out of 67), notably incorporating CSV (8 out of 67), which is not used by application or courseware repositories.

\textbf{Prompt repositories show mixed preferences for single-prompt files and multi-prompt files.} Overall, there is a nearly even split in storage conventions: 52.1\%~(48 out of 92) repositories use multi-prompt files, and 47.8\%~(44 out of 92) exclusively use single-prompt files. 
Single-prompt files offer modularity and ease of reuse but can lead to rapid growth in the number of files and increased organizational complexity. In contrast, multi-prompt files provide compactness and simpler file management but may hinder modular reuse and attribution.
Interestingly, prompt courseware repositories uniformly prefer multi-prompt files~(100\%), whereas prompt application repositories strongly favor single-prompt files~(80.0\%) since developers can easily load individual prompts directly from separate files. Prompt collection repositories show a balanced mix: 58.2\% use multi-prompt files organized sequentially without explicit content categorization, and 41.8\% adopt single-prompt files organized into directories based on topics or use-cases.

\subsection{Uneven Distribution and Prompt Duplication}


\textbf{The distribution of the number of prompts across GitHub repositories is highly skewed, with 8.7\% of GitHub repositories containing over 90\% of all collected prompts.} Figure~\ref{fig:format_and_num_prompts} (c) shows a heavily concentrated distribution of prompts, with the top eight repositories accounting for 90.8\% of the collected data. Notably, the six largest repositories are all categorized as prompt collections and account for 88.8\% of the data. Such a heavily skewed distribution raises concerns regarding representativeness for researchers mining prompts from GitHub. To address this imbalance in our own study, we do not treat the dataset as a homogeneous whole; instead, we stratify our subsequent quality analyses by repository category (i.e., Collection, Application, and Courseware). This ensures that the distinct characteristics of specialized application prompts are not overshadowed by the sheer volume of prompts in collection repositories.
Consequently, we recommend that future researchers also carefully target relevant repository categories following their specific research aims. Specifically, prompt collection repositories are better suited to large-scale analyses since they contain vast quantities of prompts~(with a median of 20 prompts). In contrast, prompt application repositories generally contain fewer but more specialized prompts~(with a median of 1 prompt), which is ideal for research that requires understanding and analysis of repositories that maintain individual prompts.

\textbf{While 10.1\%~(2,507 out of 24,800) of the analyzed prompts are exact duplicates, 38.5\%~(9,553 out of 24,800) are semantic duplicates.} We observe prompt duplication both within repositories~(internal duplication) and across multiple repositories~(external duplication). Specifically, internal exact duplication occurs in 13.0\%~(12 out of 92) of repositories, whereas internal semantic duplication occurs in 30.4\%~(28 out of 92) repositories, reflecting inefficiencies in organization and maintenance practices. Furthermore, external exact duplication occurs in 16.3\%~(15 out of 92) of repositories, while external semantic duplication occurs in 34.8\%~(32 out of 92) of repositories. These results indicate potential contamination risks and difficulties in tracking different versions of the same prompt.

%% file: sections/4_quality.tex
\section{Prompt Quality Analysis}

\begin{table*}[t]
\caption{FRE score interpretation and distribution across analyzed prompts.}
\label{tab:flesch_interpret} 
\tablefont
\begin{tabular}{llllr} 
\toprule
FRE Score & School Level & Reading Difficulty & Example Text Type & \% Prompts \\ 
\colrule
$<0$ & Highly Specialized & Extremely Difficult & Academic or legal documents & 8.1 \\
$[0,30)$ & College Graduate & Very Difficult & Technical reports, some legal documents & 29.6 \\
$[30,50)$ & College & Difficult & High school-level material & 28.1 \\
$[50,60)$ & 10th to 12th grade & Fairly Difficult & Consumer information, most web content & 14.3 \\ \colrule
$[60,70)$ & 8th and 9th grade & Plain English & Average newspapers & 12.0 \\ \colrule
$[70,80)$ & 7th grade & Fairly Easy & Magazines & 5.9 \\
$[80,90)$ & 6th grade & Easy & Popular Magazines & 1.6 \\
$[90,100)$ & 5th grade & Very Easy & Children's books & 0.4 \\
$\ge100$ & Pre-school & Extremely Easy & Picture books, basic primers & 0.1 \\ 
\botrule
\end{tabular}
\end{table*}

\begin{figure*}[t]
	\centering
	\includegraphics[width=\textwidth]{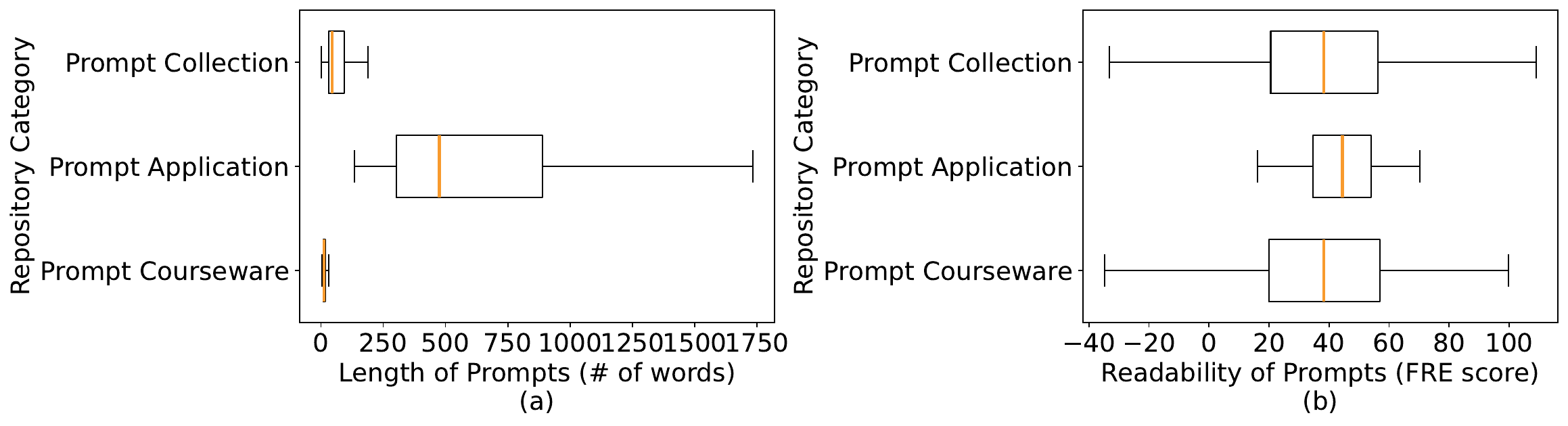}
	\caption{Comparison of prompt (a) length and (b) readability across repository categories.}
	\label{fig:prompts_length}
\end{figure*}

Like traditional software assets, prompts must adhere to certain quality attributes. This article focuses on the static quality attributes, such as readability and syntactical accuracy, to ensure usability, maintainability, and clarity for developers. In particular, prompts with poor readability can be misunderstood or incorrectly modified, leading to unintended behaviors in promptware, similar to issues observed in traditional software~\cite{piantadosi2020readability}. For example, prompts with fewer syntactic errors and higher readability correlate with better outcomes in issue resolution tasks~\cite{ehsani2025detectingpromptknowledgegaps}. Also, prompt length and the amount of background context correlate with task performance~\cite{liu2025effectspromptlengthdomainspecific}.

To systematically investigate prompt quality on GitHub, we analyze three quality metrics: prompt length, readability, and syntax correctness (i.e., spelling errors). Prompt length is measured by counting words via the regular expression \enquote{\textbackslash w+}. To measure the readability of prompts, we follow the same approach as used in previous research on LLMs~\cite{ehsani2025detectingpromptknowledgegaps,canizares2023readability}, leveraging the Flesch Reading Ease~(FRE) metric. Table~\ref{tab:flesch_interpret} shows the interpretation of FRE scores, with higher scores for texts that are easier to read. To identify statistically significant differences in prompt length and readability across repository categories~(i.e., prompt collection, application, and courseware), we perform the Mann-Whitney U test with Bonferroni correction. Given an initial significance level set at $\alpha=0.05$ and accounting for multiple comparisons (three comparisons in total), we adjust this threshold to $\alpha/3=0.017$. We also calculate Cliff's delta~$d$ effect size with 95\% confidence intervals (CI) to quantify the difference. Finally, we use the \enquote{pyspellchecker} library~\cite{pyspellchecker} to detect spelling errors for measuring syntax correctness.

\textbf{Prompts are typically short, facilitating easier review and modification, although lengths differ significantly across repository categories.} \figref{fig:prompts_length} shows that around 75\% of analyzed prompts contain fewer than 92 words, generally fitting within a single paragraph. Prompt application repositories have the highest median word count~(475 words), followed by collection repositories~(median 46 words) and then guide repositories~(median 13 words). Statistical testing reveals that prompts from application repositories are significantly longer than prompts from both collection and courseware repositories, with large effect sizes~($d=0.99$ with 95\% CI from $0.89$ to $1.10$ and $d=0.88$ with 95\% CI from $-2.44$ to $4.19$). This difference in length is mainly due to the detailed nature of application prompts, which are designed to function as natural language-based programs (i.e., promptware). For example, the prompt used by an AI tutor application~\cite{mr_ranedeer} spans over 1,700 words to specify the tutor's behaviors, functions, and configurations. In addition, prompts from collection repositories are significantly longer than those from course repositories, and they also have a large effect size~($d=0.76$ with 95\% CI from $0.56$ to $0.88$).


\textbf{Most prompts are relatively difficult to read, with 80.1\% having FRE scores below 60}, as shown in \tabref{tab:flesch_interpret}. Such a low readability prevalence may pose challenges for contributors attempting to reuse, adapt, or maintain prompts, and it may also confuse foundation models by making the intent unclear. On the other hand, 19.9\%~(4,936 out of 24,800) of the prompts achieve a FRE score of 60 or higher, indicating relative ease of reading. Unlike prompt length, we observe no significant differences in the readability of prompts across these three repository categories.

\textbf{More than half (55.2\%) of GitHub prompts contain spelling errors.} Among the 24,800 prompts analyzed, we found that 13,689 included at least one spelling mistake~(median of one typo per prompt), with substantial variation across repository categories. Prompt courseware repositories, typically intended for educational purposes, have the lowest prevalence of spelling errors~(20.5\%). Prompts from collection repositories show a moderate prevalence~(56.0\%), while prompts from application repositories demonstrate the highest prevalence of spelling errors~(96.7\%). This high error rate in application repositories is likely due to the difficulty of maintaining longer prompts and limited quality assurance mechanisms.

%% file: sections/6_threats.tex
\section{Threats to Validity}

\textbf{Internal Validity.} Our definition of prompt quality relies on static metrics such as readability, length, and spelling correctness rather than dynamic execution metrics. We acknowledge that these metrics do not guarantee the correctness or effectiveness of the LLM's output. While correlating prompt characteristics with LLM performance is a valuable direction, it lies outside the scope of this study, which focuses on the maintainability and management of prompts as software assets.

\textbf{External Validity.} Our findings are based on a dataset of open-source prompts hosted on GitHub. We filtered for repositories with at least 10 stars and containing English-language prompts to ensure relevance and quality. This sampling strategy excludes smaller, niche, or non-English repositories. Consequently, our findings may not generalize to early-stage projects that have not yet gained popularity (fewer than 10 stars) or to prompts written in other languages, which may exhibit different structural or linguistic characteristics. In addition, our focus on GitHub may not fully capture practices in other ecosystems, such as Hugging Face or private commercial prompt stores.

A potential limitation is the restriction of the dataset to prompts collected through December 2023, given the evolution of prompt engineering. Expanding the full dataset is infeasible due to the significant manual effort required to extract prompts from unstructured repositories without standardized file extensions. To address this, we conducted a validation study on 20 randomly sampled repositories (10 repositories each year) from 2024 and 2025 (1,787 prompts). The results confirm consistency with the original dataset. First, Markdown remains the dominant format, appearing in 85\% of the sampled repositories. Second, the distribution of prompt length across categories mirrors previous patterns. Finally, quality issues persist and have worsened, with a 72.4\% spelling error rate in the newer sample compared to 55.2\% in the main dataset. This indicates that the management and quality assurance challenges identified in the 2021 to 2023 period remain critical in the current landscape.

%% file: sections/5_lessons.tex
\section{Learned Lessons}

\textbf{Standardize prompt formats and organizational practices.}
Although Markdown is the dominant format for storing prompts, we observe notable inconsistencies regarding file organization~(e.g., single-prompt vs. multi-prompt files) and formatting conventions~(e.g., prompt structure, author attribution, context documentation). Such inconsistencies pose challenges for efficient reuse, attribution, and automated mining of prompts. To address these challenges, communities and industry stakeholders should establish clear guidelines and standardized formats designed specifically for prompt management using code repositories. These standards should emphasize machine-readable structured metadata (e.g., authorship, intended use-cases, and creation context) alongside human-readable documentation.

\textbf{Improve prompt discoverability and reuse by structuring and categorizing prompts.}
Our topic analysis reveals diverse use cases for prompts, however, 52.1\% of prompt repositories store multiple prompts in a single file without explicit categorization or tagging. This practice makes it difficult for developers to search and reuse prompts for particular tasks. We suggest that developers and maintainers use structured directories or file formats (e.g., CSV) that organize prompts based on their use cases.

\textbf{Mitigate prompt duplication by integrating automated duplicate detection tools.}
Prompt duplication, both within and between repositories, mirrors traditional software duplication issues~\cite{rattam2013clone}. Such duplication risks maintenance inefficiencies, error propagation, and ambiguity in prompts' original authorship or modification histories. 
Developers should integrate automated duplicate detection tools into their workflows, regularly auditing prompts to reduce redundancy and comprehensively document provenance, thereby reducing duplication-related risks.

\textbf{Integrate automated prompt quality assessment into repository maintenance, analogous to CI/CD pipelines.}
Our analysis highlights substantial variability in prompt readability and spelling accuracy across repositories, indicating insufficient quality control mechanisms. Borrowing from traditional software best practices, continuous integration and continuous deployment~(CI/CD) could be implemented to regularly monitor prompt quality. Integrating automated tools for readability assessment, syntax verification~(e.g., spell-checking), and metadata validation into routine repository maintenance can benefit developers, promoting trust and encouraging contributor engagement.
For instance, a repository could configure a GitHub Action to automatically trigger a spell checker whenever a prompt file is modified in a pull request. If a contributor submits a prompt with errors (e.g., \enquote{generte} instead of \enquote{generate}), the CI pipeline would flag the issue and block the merge until resolved. This automated gatekeeping ensures that only high-quality prompts enter the codebase.